\documentclass[11pt]{article}
\usepackage{moriond,epsfig}

\def\be{\begin{equation}}
\def\ee{\end{equation}}
\def\bea{\begin{eqnarray}}
\def\eea{\end{eqnarray}}

\def\MW{m_\mathrm{W}}
\def\MZ{m_\mathrm{Z}}
\def\MH{m_\mathrm{H}}
\def\Mt{m_\mathrm{t}}
\def\GW{\Gamma_\mathrm{W}}
\def\MeV{\rm Me\kern -0.1em V}
\def\GeV{\rm Ge\kern -0.1em V}
\begin{document}
\vspace*{4cm}
\title{W MASS AT LEP AND STANDARD MODEL FITS}

\author{ S. ROTH }

\address{III.~Physikalisches Insitut, RWTH Aachen University,\\
         D-52056 Aachen, Germany}

\maketitle
\abstracts{
The W-mass measurements from LEP and the results of a global fit
of the Standard Model parameters to the electroweak data are presented.
Comprehensive studies of experimental systematic effects allowed a 
measurement of the W mass with an accuracy of better than half a permill.
Especially the recent improvements in the LEP energy calibration, 
the modelling of the hadronisation process and the understanding 
of Bose-Einstein correlations and colour reconnection effects are discussed.
The fit of the Standard Model parameters to all electroweak measurements
verifies the self-consistency of the theory.
The combination of all electroweak data yield information on the mass of 
the still undiscovered Higgs boson, $\MH$.
}

\section{Measurement of the W mass at LEP}

The mass, $\MW$, and the total decay width, $\GW$, are fundamental 
properties of the W boson.
Comparing $\MW$ to its predictions derived from other electroweak 
parameters provides a stringent test of the Standard Model at the 
level of quantum corrections.
In addition to the measurement of the Z mass and the weak mixing angle 
performed at LEP~1, an accurate measurement of the W mass is mandatory
for this test.
The precision of the direct measurement of the W mass has to
compete with the $23~\MeV$ accuracy on $\MW$ when it is derived 
indirectly from the electroweak precision data.

The results presented here are obtained 
using about 40,000 W-pair events recorded by the LEP experiments 
at centre-of-mass energies, $161~\GeV < \sqrt{s} < 209~\GeV$.
The cross section of W-pair production near threshold is sensitive to $\MW$.
Here, $\MW$ is derived from the measurement of the total cross section 
of W-pair production~\cite{wmass-threshold}.
At higher centre-of-mass energies, well above the kinematic threshold,
the W-pair events are directly reconstructed and the invariant mass
spectra of the W-boson decay products are exploited~\cite{wmass-direct}.

In the following the current status of the results on the W mass
from LEP is presented.
It is based on final results from the experiments ALEPH, L3 and OPAL
and preliminary numbers from DELPHI.
In all analyses the final calibration of the LEP beam energy is used.
Together with the improvemed understanding of colour reconnection and
Bose-Einstein effects in the fully-hadronic final state a significant 
reduction of the W-mass uncertainty with respect to the last year was achieved.
The combination of the results from the four experiments yields
\begin{eqnarray}
   \MW & = & 80.388 \pm 0.026 \pm 0.024~\GeV \; ,
\end{eqnarray}
where the first uncertainty is statical and the second is systematic.

\subsection{LEP beam energy calibration}

The calibration of the LEP beam energy is based on the resonant 
spin depolarisation (RDP) technique available at beam energies of 
$41~\GeV < E_\mathrm{beam} < 61~\GeV$.
Unfortunately this method can not be used for the beam energies of the 
physics runs at LEP~2, above $60~\GeV$.
Therefore, the measurements of resonant depolarisation
made at low energies are used instead, to calibrate an energy measurement
which is based on B-field measurements accomplished with 16 NMR probes 
situated in selected bending dipoles~\cite{lep-ebeam}.
They were read out during physics runs as well as during the procedure of
resonant depolarisation.
The beam energies for the physics running in the regime
$80~\GeV < E_\mathrm{beam} < 104~\GeV$ were derived from
the NMR model extrapolating the results of the RDP technique to the higher
energies.

\begin{figure}
\centering
\includegraphics[height=0.25\textheight]{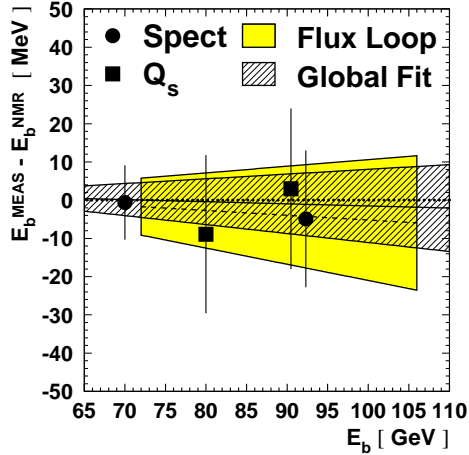}
\caption{
    Differences between the results from the NMR model and the alternative
    methods, using the magnetic spectrometer, the flux loop coils and the
    synchrotron tunes.
    The systematic uncertainty assigned to the LEP energy measurement due
    to this comparison is depicted as shaded area.
}
\label{fig:lep-ebeam}
\end{figure}

The systematic uncertainty of the NMR model was derived using three
independent measures of the LEP beam energy:
the synchrotron tune, $Q_s$, of the LEP storage ring,
the flux-loop, a sequence of cable loops installed into each of the
bending magnets and sensing the change of the magnetic flux during the ramp
of the B field, and a magnetic spectrometer installed in 1999 and used 
during the run of the year 2000.
The relative differences between the result obtained from the NMR model
and the alternative measurements are shown in Figure~\ref{fig:lep-ebeam}.

The comparison of the alternative methods with the default NMR
measurement allows to estimate the systematic uncertainty of the 
LEP beam energy.
This uncertainty increases linearly with the distance in energy 
to the calibration points where the precise energy calibration 
using the depolarisation method was performed.
Finally uncertainties of $10 - 20~\MeV$ are assigned to $E_\mathrm{beam}$
depending on the running period.
The contribution of the beam energy calibration to the uncertainty on
the final W mass accounts to $9~\MeV$.

\subsection{Hadronisation}

After the generation of the four-fermion state the quark pairs
are subject to a Monte-Carlo program modelling the hadronisation process.
Three different schemes, implemented in the programs Pythia, Herwig and 
Ariadne, are widely used for this purpose.
QCD studies of the LEP~1 data were not able to decide between the predictions
of the three programs.

Therefore, systematic effects due to modelling of the hadronisation process 
are determined by comparing the result of three different mass fits using
Monte-Carlo events simulated with the programs Pythia, Herwig and Ariadne, 
respectively.
Here, identical events at the level of the four-fermion state were subject 
to all three hadronisation programs.
In addition, DELPHI and OPAL compared Pythia Monte Carlo samples generated 
with various QCD parameters, for example the hadronisation scale 
$\Lambda_\mathrm{QCD}$ and the shower parameter $\sigma_q$, 
varied with respect to the standard tuning of the Pythia generator.

\begin{figure}
\centering
\includegraphics[width=0.8\textwidth]{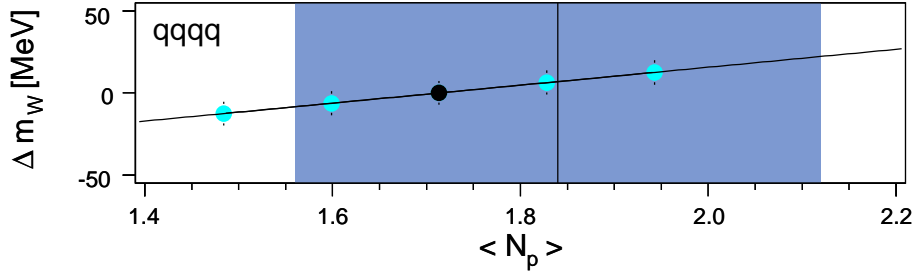}
\caption{
    Change of $\MW$ in the L3 analysis when re-weighting Monte Carlo events
    with respect to the proton multiplicity of $qqqq$ events.
    The full circle shows the default Pythia value whereas the vertical line
    shows the multiplicity measured.
}
\label{fig:proton}
\end{figure}

In the L3 analysis the Pythia Monte Carlo events are re-weighted such that
the mean number of charged kaons and the mean number of protons agree with the
measurement~\cite{kaon-proton}.
The resulting shift of the W mass is shown in Figure~\ref{fig:proton}
for one example.
The extracted W mass depends linearly on the number of kaons and protons.
Due to this linearity the uncertainty of the measured kaon and proton
multiplicities can be translated into an uncertainty on the W mass.

\subsection{Bose-Einstein effects}

Bose-Einstein correlations are caused by interference effects between
identical bosons which are close to each other in phase space.
They manifest themselves as an enhanced production of 
identical bosons, mainly pions, at small four-momentum difference.
For quantitative studies of Bose-Einstein effects in W-pair events and for the
estimation of possible mass biases the Monte-Carlo model 
BE32 implemented in the program code LUBOEI~\cite{luboei} from 
L\"onnblad and Sj\"ostrand is widely used.

All four LEP experiments published their final results on Bose-Einstein 
correlations in W-pair events~\cite{be-lep}.
Typically one million of like-sign particle pairs are selected in the 
fully-hadronic and about 200,000 in the semi-hadronic channel,
using all data of one LEP experiment.
A comparison of Bose-Einstein correlations in fully-hadronic W-pair 
events ($qqqq$) with those in semi-hadronic events ($qq\ell\nu$) serves 
as a probe to study the inter-W Bose-Einstein correlations.

\begin{figure}
\includegraphics[height=0.25\textheight]{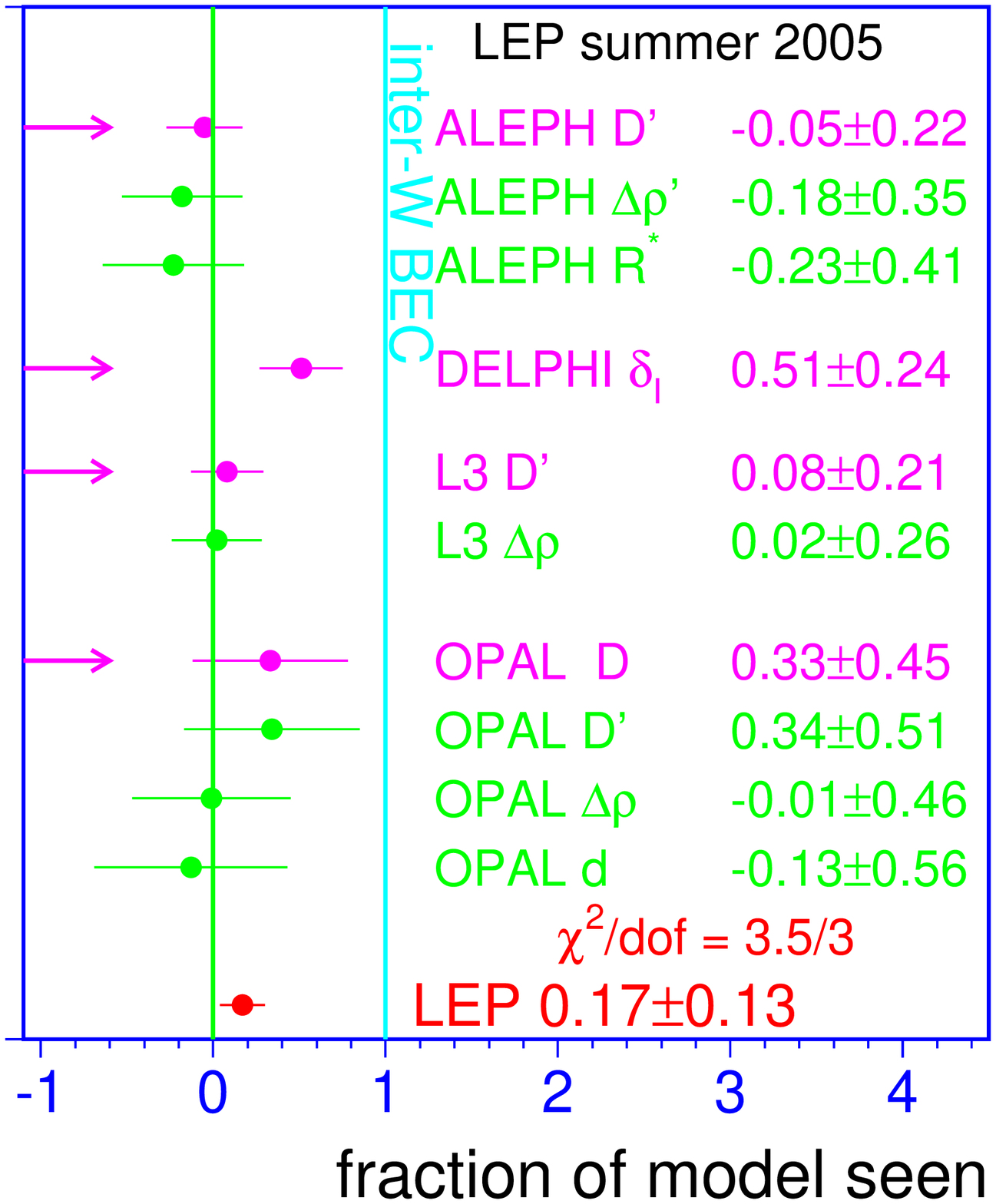}
\hfill
\includegraphics[height=0.25\textheight]{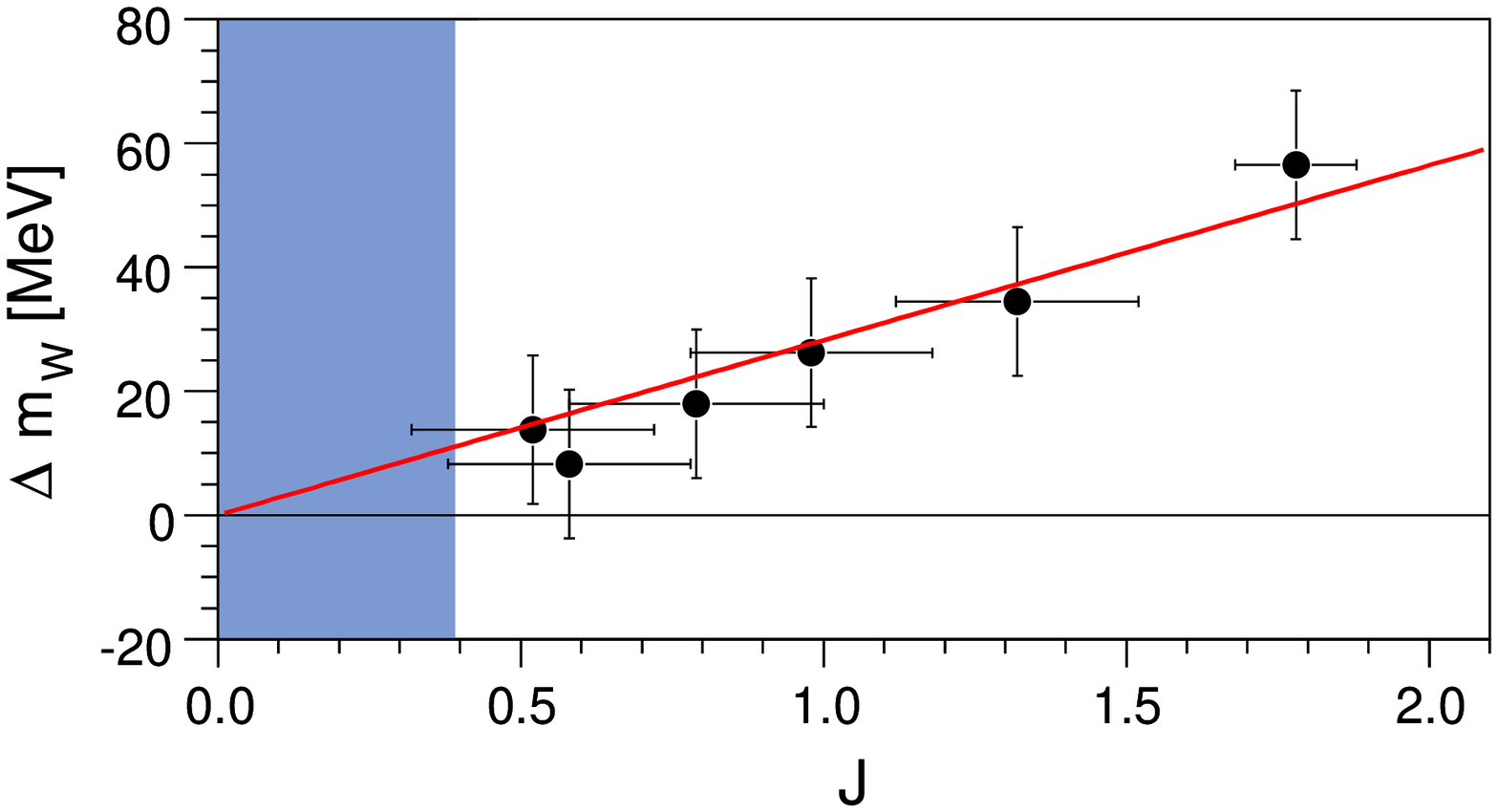}
\caption{
   Left: Measured Bose-Einstein correlations given in terms of
   a relative fraction to the size expected from the LUBOEI model
   including inter-W correlations.
   Right: Linearity of the W-mass shift with respect to the Bose-Einstein 
   observable $J$.
}
\label{fig:be}
\end{figure}

A combination of the results of the four LEP experiments has been 
performed by averaging the results of various analyses with different 
estimators for the size of Bose-Einstein correlations with respect to 
the specific model under study.
Figure~\ref{fig:be} shows the measured size of correlations as 
a relative fraction of the LUBOEI model including inter-W correlations.
Combining the individual results gives an average fraction of
$0.17 \pm 0.13$.
This means that the data prefer only little inter-W correlations,
at most at the level of about one third of the LUBOEI model.

For the L3 W-mass analysis several Monte Carlo samples with various
strengths of inter-W Bose-Einstein correlations, but equal strength
of intra-W correlations were subject to the mass extraction procedure.
The Bose-Einstein observable $J$ is determined for each of these samples.
A linear dependence of the W-mass shift with respect to $J$ is
obtained, as presented in Figure~\ref{fig:be}.

The measurement of the W mass is based on a Monte-Carlo simulation
using the LUBOEI BE32 model without inter-W correlations.
To estimate the systematic uncertainty due to Bose-Einstein effects
the fit is also performed using a simulation with full inter-W correlations.
According to the Bose-Einstein measurement currently 30\% of the difference 
between both results are taken as systematic uncertainty.
In total the Bose-Einstein effect contributes with $13~\MeV$ to the
uncertainty on the W mass in the $qqqq$ channel.

\subsection{Colour reconnection}

In the fully-hadronic final state of W-pair production both W bosons decay 
into a quark-antiquark pair which subsequently hadronise into jets.
The spatial extension of the hadronisation process is given by the
range of the strong interaction of about 1~fm which is about ten times
larger than the decay length of the W boson leading to a significant
space-time overlap of the two colour strings.
Due to this overlap the W bosons may not hadronise independently, 
i.e.\ a re-arrangement of the colour flow is possible.
This effect is called colour reconnection.

The effect of colour reconnection in the non-perturbative phase can
be modelled with the SK models developed by Sj\"ostrand and 
Khoze~\cite{sk-model} based on the string fragmentation model implemented 
in Pythia.
The colour strings are assumed to have a finite width of about
1~fm, the typical range of the strong interaction.
Therefore the two strings originating from the two hadronically
decaying W bosons can exhibit a non-vanishing space-time volume
over which they overlap.
In the SK-I model the reconnection probability between the two strings
is proportional to this space-time overlap multiplied by a free parameter,
$k_I$.

Detailed searches for colour reconnection in hadronic decays of W pairs
were performed using the event particle flow method~\cite{cr-lep}.
In this analysis, the four jets are grouped into the two pairs originating
from the decay of a W boson and the particle flow in four different jet-jet
regions are studied: two regions between jets with the same parent W boson
(intra-W) and two in which the parents differ (inter-W).
If colour reconnection between the jets of different W decays is present,
the particle flow between the jets would be changed.
More particles are expected to be emitted between jets from different W bosons.
To quantify this effect, the particle flow in between jets coming from the
same W (intra-W region) is divided by the particle flow in the regions 
between the W bosons (inter-W region).
The measured ratio is then compared to the predictions from Pythia without
colour reconnection and the prediction from the SK-I model.

\begin{figure}
  \includegraphics[height=0.25\textheight]{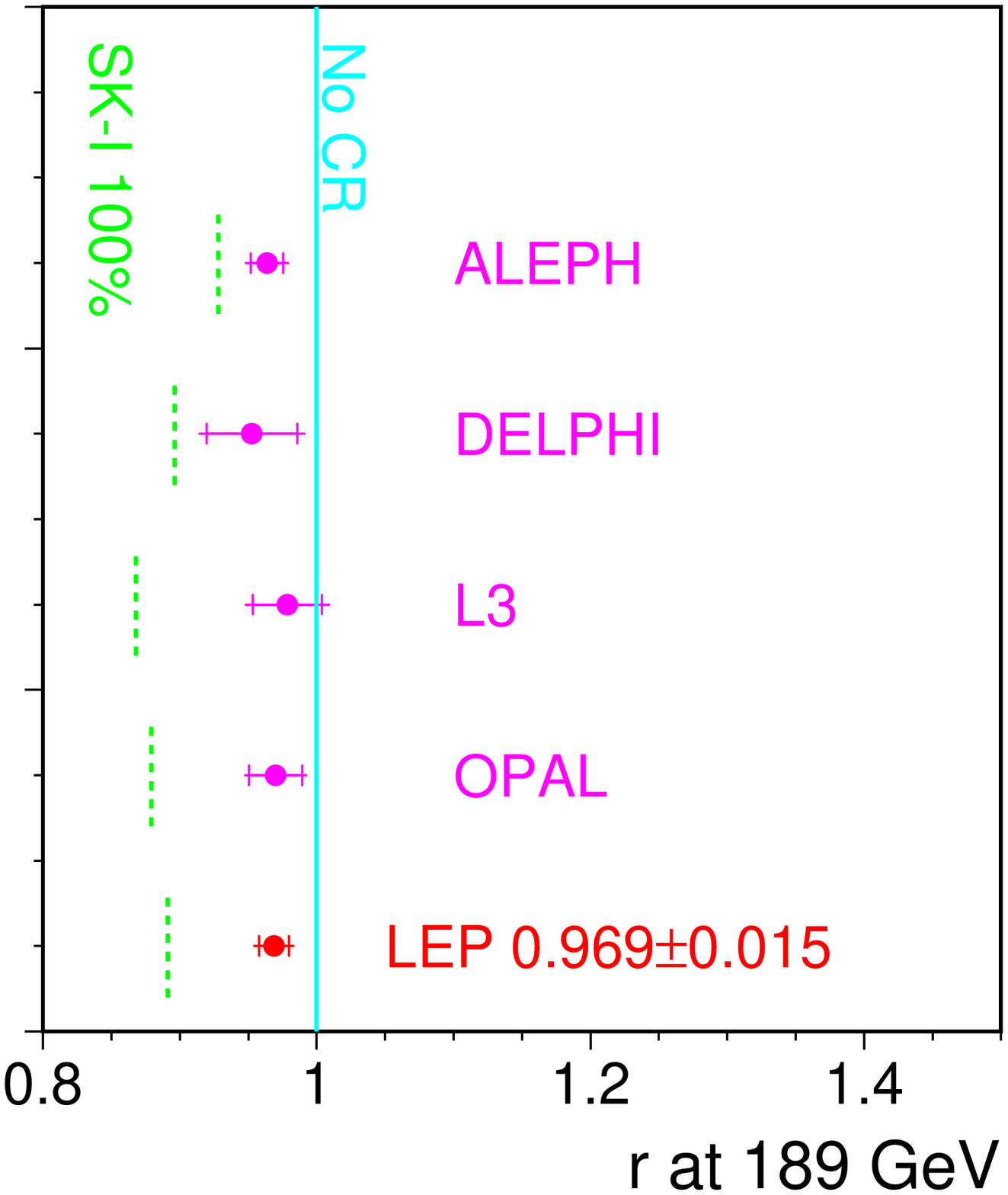}
  \hfill
  \includegraphics[height=0.25\textheight]{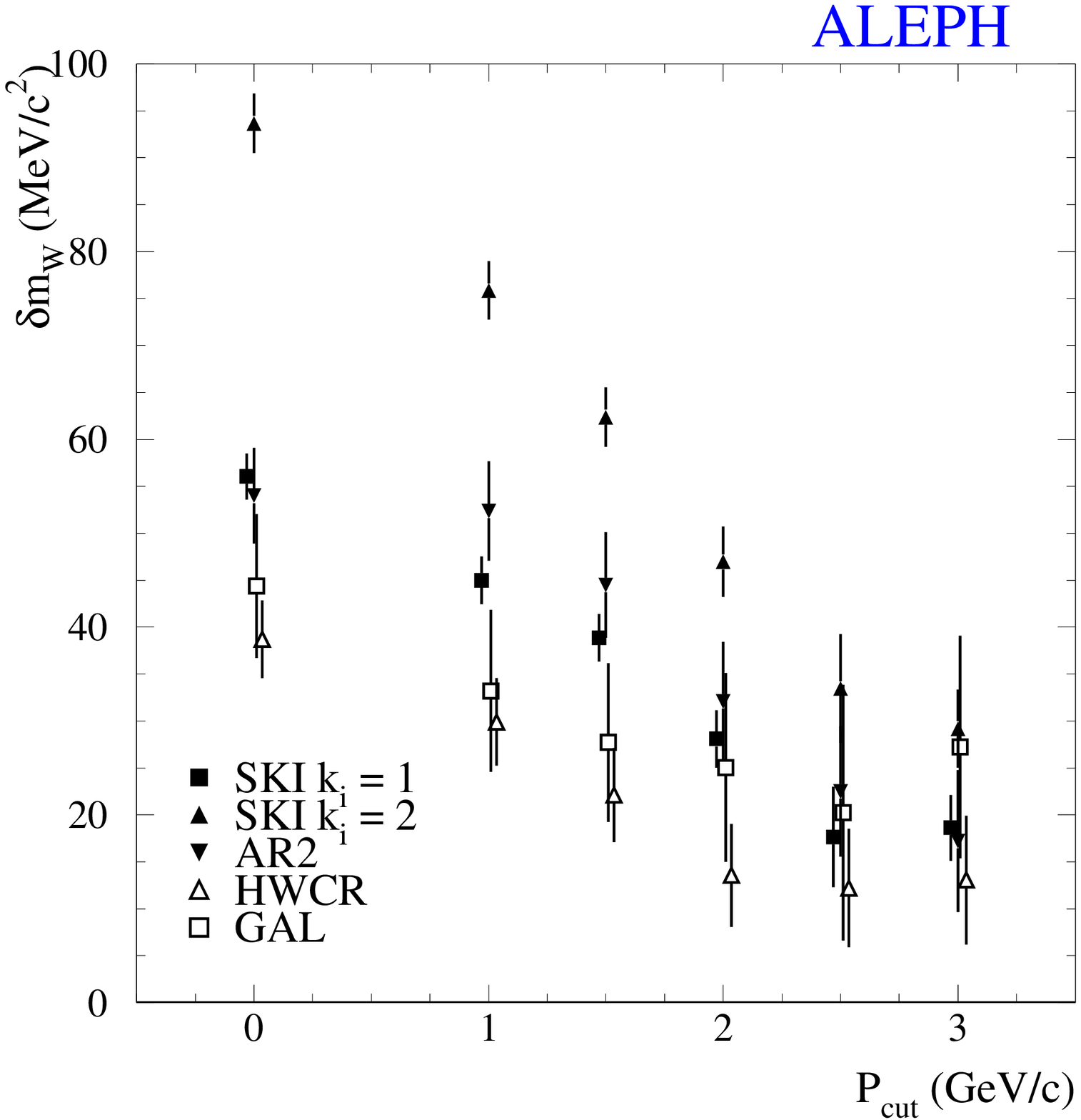}
  \caption{
  Left: Preliminary result on the combination of the particle flow results
  of the four LEP experiments;
  the predicted value of $r$ for the SK-I model with full colour reconnection
  is shown for each experiment by the dashed lines.
  Right: Shift of the W mass between the standard Monte Carlo and
  various colour reconnection models as function of the
  cut on the minimum particle momentum, $p_\mathrm{cut}$.
  }
\label{fig:cr}
\end{figure}

In Figure~\ref{fig:cr} the results of the four analyses of
the LEP experiments are compared to each other and to the prediction
of the SK-I model with full colour reconnection.
The combined average slightly prefers a non-vanishing contribution
from colour reconnection.
This is mainly caused by the ALEPH result.
An upper limit at 68\% confidence level is set at $k_\mathrm{I}=2.1$.
The data disfavour the extreme version of the SK-I model with full
colour reconnection by 5.2 standard deviations.

Using a cone algorithm for jet clustering lowers the sensitivity
to colour reconnection effects, as the analysis will be unaffected by the 
inter-jet regions where the influence of the CR is largest.
Alternatively, removing clusters below a certain momentum cut rejects
particles predominantly produced during the non-perturbative phase of the
hadronisation process where the colour reconnection effects take place.
The shift of the W mass observed between the standard Monte Carlo and 
various colour reconnection models is shown in
Figure~\ref{fig:cr} for the ALEPH analysis.
This mass shift is evaluated for variations of the jet reconstruction
where a cut on the minimum particle momentum, $p_\mathrm{cut}$,
was applied.
For stronger cuts on $p_\mathrm{cut}$ the W mass shifts 
due to colour reconnection are significantly reduced.

In their final analyses all LEP experiments decided to limit the
effects of colour reconnection by introducing such a cut in the 
jet reconstruction.
Using the preliminary upper limit on $k_\mathrm{I}$ presented above
the uncertainty from colour reconnection contributes with $31~\MeV$
to the uncertainty of the W mass in the $qqqq$ channel.
Further improvement is expected when the final colour reconnection results
of all four LEP experiments will become available.

\section{Standard Model fits}

Electroweak radiative corrections have been calculated up to the two-loop 
level, but their accuracy is limited by the experimental uncertainty
on the mass of the top quark and by the ignorance on the mass of the 
Higgs boson.
A test of the quantum structure of the Standard Model therefore requires
a precise knowledge of the top quark mass, since the radiative corrections
depend quadratically on this parameter.
The electroweak radiative corrections include a term proportional to the 
logarithm of the Higgs mass.
Assuming the validity of the Standard Model one can try to extract 
this term and hence the Higgs mass from a global fit to all electroweak 
observables.

The mass of the top quark is measured by the Tevatron experiments CDF and D0
analysing events of the reaction 
$\mathrm{p} \bar\mathrm{p} \to \mathrm{t} \bar\mathrm{t} X 
\to \mathrm{b} \bar\mathrm{b} \mathrm{W}^+ \mathrm{W}^- X$.
The published Run-I results have been combined with the most recent 
Run-II measurements representing a data set of 750~pb${}^{-1}$ 
in total~\cite{mtop}:
\[
   \Mt = 172.5 \pm 2.3 \; \GeV \; .
\]

The fit results presented here are performed within the context of the 
LEP and SLD electroweak working group.
The details of the combination of the electroweak data and the fit of
the Standard Model parameters are described in Reference~\cite{ewwg}.
The semi-analytical program ZFITTER~\cite{zfitter} is used to calculate 
the Standard Model predictions including its higher order corrections.
The complete fermionic and bosonic two-loop corrections to $\MW$ 
have been calculated recently~\cite{twoloop}.
The analytical formulas obtained were parametrised as functions of the 
parameters $\MH$, $\Mt$, $\MZ$, $\Delta \alpha$ and $\alpha_\mathrm{s}$
and implemented in ZFITTER.

The most general electroweak fit includes all Z peak data from LEP~1 and SLD,
the W mass, $\MW$, from LEP~2 and Tevatron, the top mass, $\Mt$,
measured at the Tevatron and $\Delta \alpha^{(5)}_\mathrm{had}$.
This fit yields a $\chi^2/\mathrm{d.o.f.} = 17.5 / 13$ which corresponds 
to a fit probability of 18\%.
This shows that the electroweak measurements are internally consistent 
and agree with the Standard Model prediction.
The radiative corrections are needed to describe the data which represents 
a test of the Standard Model at the quantum-loop level.

\begin{figure}
  \includegraphics[height=0.25\textheight]{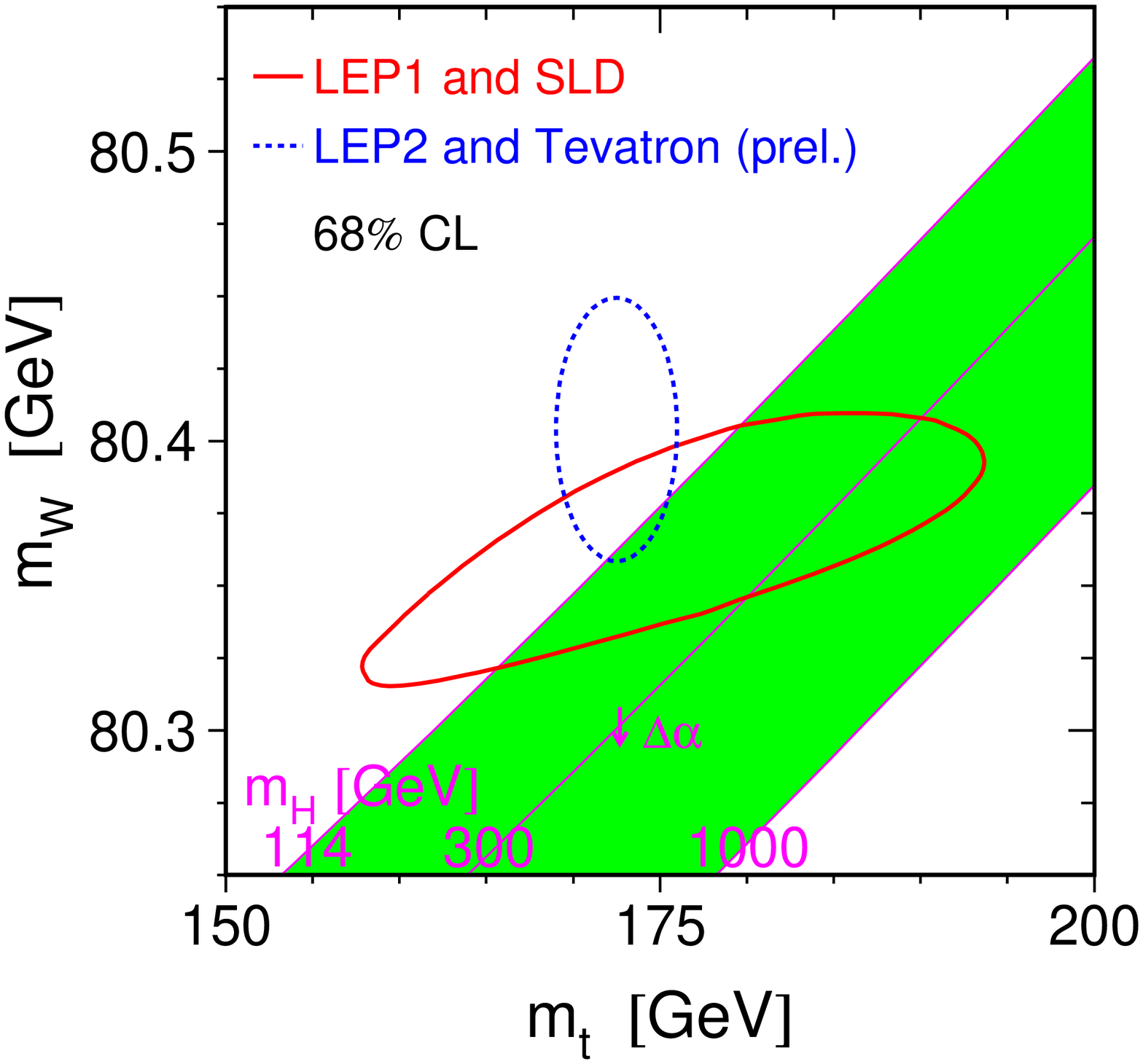}
  \hfill
  \includegraphics[height=0.25\textheight]{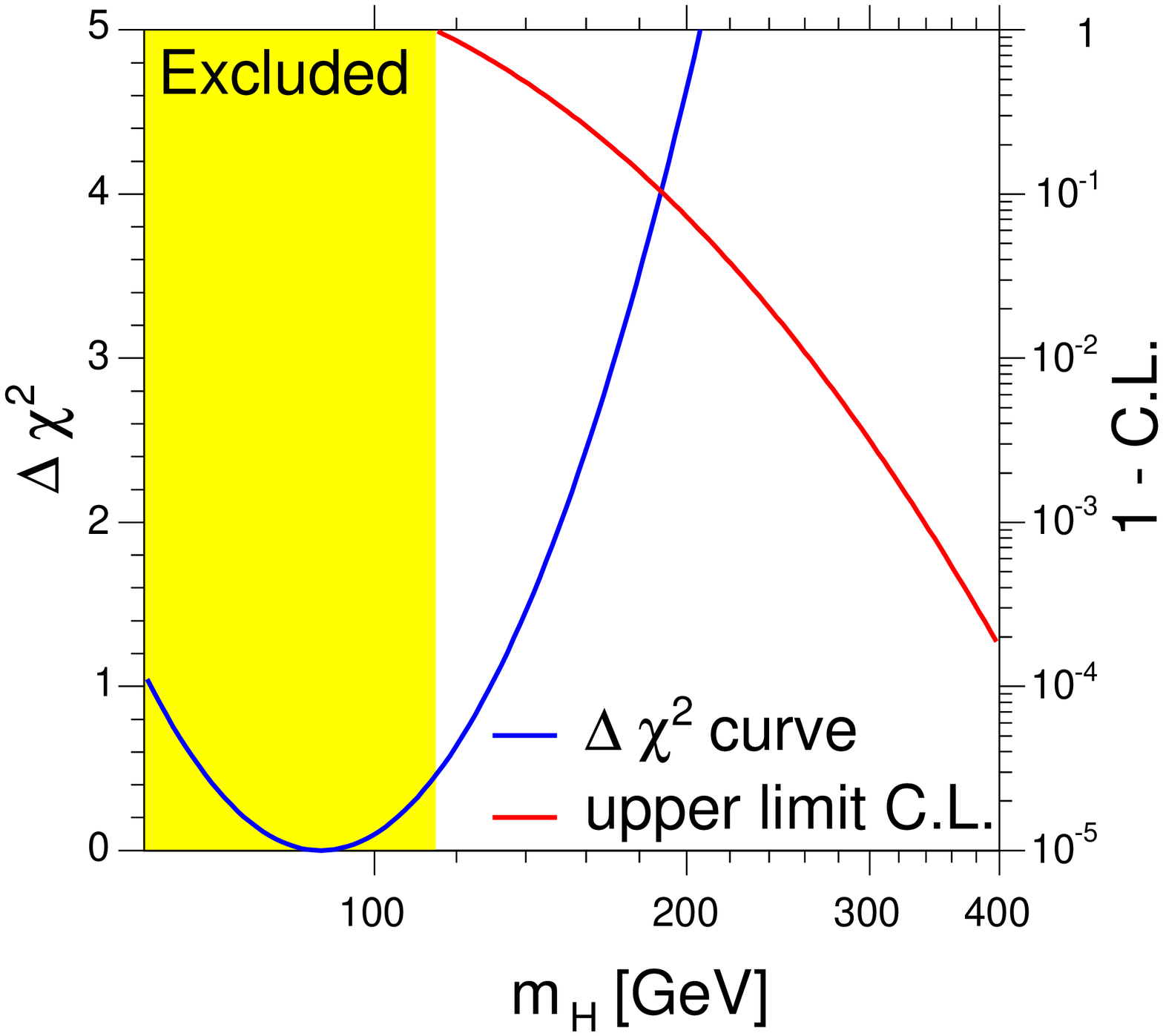}
  \caption{
  Left: Comparison of direct mass measurements and indirect measurements 
    using electroweak precision data;
    also shown is the theory prediction for various values of $\MH$.
  Right: The $\Delta \chi^2$ curve of the Standard Model fit;
    in addition the confidence level (C.L.) as a function of the
    upper limit on $\MH$ is shown. 
  }
\label{fig:sm-fit}
\end{figure}

In Figure~\ref{fig:sm-fit} the allowed region of all LEP~1 and SLD 
electroweak data is shown as a contour in the $\MW$ vs. $m_\mathrm{t}$ 
plane containing a probability of 68\%.
The direct measurements of the W boson mass and the top quark mass
are also indicated.
The Standard Model prediction derived from the precision parameter 
$G_\mathrm{F}$ is plotted for 
various Higgs masses within $114~\GeV < \MH < 1000~\GeV$.
Both the indirect and the direct measurements prefer a low Higgs mass.

The fit of the Standard Model prediction to all electroweak 
data with the Higgs mass as the only free parameter results 
in a $\chi^2$ curve as shown in Figure~\ref{fig:sm-fit}.
It predicts the mass of the Higgs boson to be $89^{+42}_{-30}~\GeV$
which is consistent with the direct searches for the Higgs boson
excluding masses below $114.4~\GeV$.
After integrating the probability density with respect to $\MH$ and 
setting the total probability of the Higgs mass above $114.4~\GeV$ to unity, 
a curve representing the confidence level on an upper limit on $\MH$ 
is obtained.
At 95\% C.L.\ an upper bound on the Higgs mass of $207~\GeV$ is set.

\end{document}